\documentclass[11pt]{iopart}

\usepackage{iopams}
\usepackage{amssymb}
\usepackage{epsfig}
\usepackage{epstopdf}
\usepackage{graphicx}
\usepackage{soul,color}
\usepackage[sort&compress,numbers]{natbib}
\usepackage[breaklinks=true,colorlinks=true,linkcolor=blue,urlcolor=blue,citecolor=blue]{hyperref}
\usepackage{dcolumn}
\begin{document}
\title[]{Post-growth purification of Co nanostructures prepared by focused electron beam induced deposition}

\author{E Begun$^1$, O V Dobrovolskiy$^{1,2}$, M Kompaniiets$^1$, R Sachser$^1$, Ch Gspan$^3$, H Plank$^{3,4}$, and M Huth$^1$}
\address{$^1$ Physikalisches Institut, Goethe University, 60438 Frankfurt am Main, Germany}
\address{$^2$ Physics Department, V. Karazin National University, 61077 Kharkiv, Ukraine}
\address{$^3$ Graz Centre for Electron Microscopy, 8010 Graz, Austria}
\address{$^4$ Institute for Electron Microscopy and Nanoanalysis, TU Graz, 8010 Graz, Austria}
\eads{\mailto{Dobrovolskiy@Physik.uni-frankfurt.de}}
\begin{abstract}
In the majority of cases nanostructures prepared by focused electron beam induced deposition (FEBID) employing an organometallic precursor contain predominantly carbon-based ligand dissociation products. This is unfortunate with regard to using this high-resolution direct-write approach for the preparation of nanostructures for various fields, such as mesoscopic physics, micromagnetism, electronic correlations, spin-dependent transport and numerous applications. Here we present an in-situ cleaning approach to obtain pure Co-FEBID nanostructures. The purification procedure lies in the exposure of heated samples to a H$_2$ atmosphere in conjunction with the irradiation by low-energy electrons. The key finding is that the combination of annealing at $300^\circ$C, H$_2$ exposure and electron irradiation leads to compact, carbon- and oxygen free Co layers down to a thickness of about 20\,nm starting from as-deposited Co-FEBID structures. In addition to this, in temperature-dependent electrical resistance measurements on post-processed samples we find a typical metallic behavior. In low-temperature magneto-resistance and Hall effect measurements we observe ferromagnetic behavior.
\end{abstract}

\maketitle

\tableofcontents

\section{Introduction}
The local deposition of materials by means of focused electron beam induced deposition (FEBID) represents a versatile approach for the fabrication of functional nanostructures. The proven applications of FEBID range from photomask repair~\cite{Lia05vst}, fabrication of nanowires~\cite{Fer13scr} and nanopores~\cite{Dan06lan} over magnetic~\cite{Gab10nan,Fer09apl} and tunable strain sensors~\cite{Sch10sen} to artificial pinning structures for Abrikosov vortices in superconductors~\cite{Dob10sst,Dob11pcs}, heterostructures for proximity-induced superconductivity~\cite{Kom14jap} and direct-write superconductors~\cite{Win14apl}. However, a long-standing problem lies in that for most of the organo-metallic precursors the low efficiency of the process for decomposing the precursor gas molecules gives rise to a rather large abundance of C in the deposits with inclusions of oxygen, whereas the metal percentage is low.

In general, FEBID structures prepared with organometallic precursors are nano-granular metals. In these, metallic grains of a few nm in diameter are embedded in a carbonaceous matrix, due to an incomplete dissociation of the precursor molecules. Owing to the sensitivity of the matrix to post-processing treatments, the compositional, structural, and, hence, electrical~\cite{Sac11prl,Por11jap} and magnetic~\cite{Ber10cie} properties of metal-based layers fabricated by FEBID can be substantially modified either in-situ or ex-situ. Exemplary purification treatments of samples include annealing in reactive gases~\cite{Bot09nan}, electron irradiation~\cite{Sac11prl,Por11jap}, or a combination of both~\cite{Meh13nan,Pla14ami,Gei14jpc}.

Thus, for Pt-based deposits, using the precursor Me$_3$CpMePt, strategies have been developed to obtain clean metal structure by in-situ post-growth treatments using O$_2$ or H$_2$O as reactive gases~\cite{Gei14jpc,Meh13nan,Pla14ami,Sac14ami}. For Co- and Fe-based deposits, using the precursors Co$_2$(CO)$_8$ and Fe(CO)$_5$, respectively, several reports have been given stating metal contents well above $80$~at.\% in as-grown samples~\cite{Lau02vst,Utk05aem,Nik12apl}. By careful optimization even up to $95$~at.\% metal purity was reported for Co structures in very few cases~\cite{Ser11acs,Fer09jpd,Cor10men}. However, such high metal contents are by no means obtained routinely, even if special care is taken to work under optimized high-vacuum conditions, such as H$_2$O removal from residual gases by Meissner traps and pre-growth plasma cleaning of the scanning electron microscope (SEM) chamber.

Here we take a different approach and present an evaluation of different in-situ post-growth purification methods for Co-based FEBID structures employing H$_2$, electron-beam irradiation and combinations of both at elevated substrate temperatures. We find that pure Co is obtained in a near-surface layer of approximately 20\,nm thickness under combined irradiation treatment and H$_2$ exposure. Temperature-dependent resistivity, magnetoresistance and Hall effect measurements corroborate the metallic nature of the purified deposits.
\section{Sample preparation and characterization}
\subsection{Preparations and pre-treatments}
Co growth, post-growth processing and imaging experiments were carried out in a dual-beam high-resolution SEM~(FEI, Nova NanoLab 600) with a Schottky electron emitter. The SEM was equipped with a FEI-made automatic gas injection system for FEBID of Co with Co$_2$(CO)$_8$ as precursor gas and with a second home-made gas injection system for feeding H$_2$ into the SEM chamber. For heating of the samples inside the SEM, we used a heatable home-made stage with a ceramic carrier chip, whose details can be found elsewhere~\cite{Sac14ami}. As substrates we used Si/SiO$_2$(10~nm)/Si$_3$N$_4$(100~nm) with Cr/Au contacts of $5/40$~nm thickness prepared by photolithography in conjunction with lift-off. Before use, the substrates were chemically cleaned with acetone, isopropanol and distilled water in an ultrasound bath. Prior to the FEBID process, the following procedures have been undertaken with the purpose of optimizing the high-vacuum conditions and minimizing the contamination of both, the SEM chamber and the substrate.

First, to passivate the inner surfaces owing to possible contaminations, the SEM chamber was cleaned for $4$ hours with a plasma source using ambient air~\cite{Mut12bjn}. After $50$ hours of pumping the base pressure amounted to $6\times10^{-6}$~mbar.

\begin{figure}
\centering
   \includegraphics[width=0.5\textwidth]{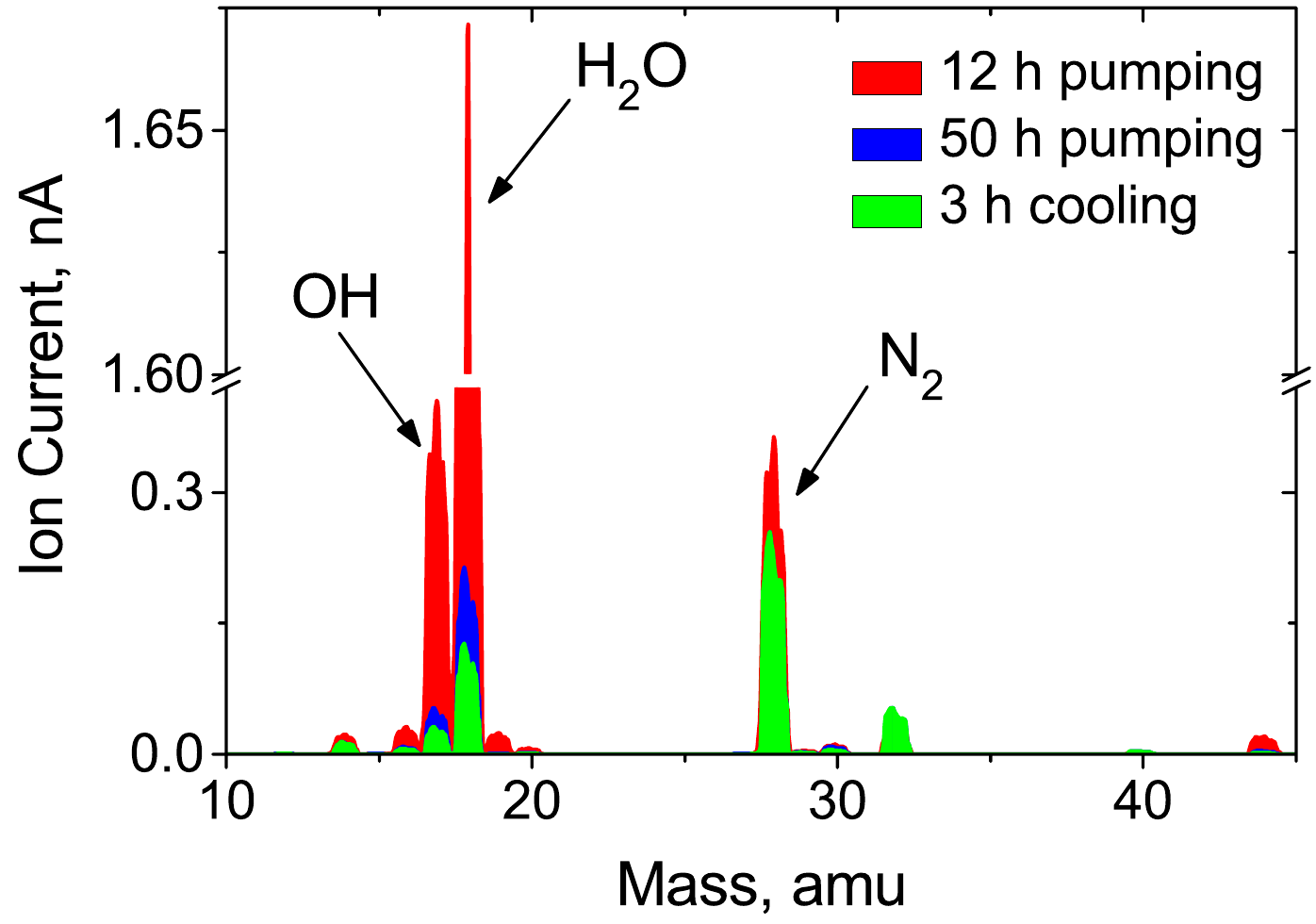}
    \caption[]
    {Mass spectrograms for the chamber atmosphere after different pre-treatments, as indicated.}
   \label{fMS}
\end{figure}

Next, the carrier chip was degassed and kept heated at $300^\circ$C in the SEM for $2$~hours, to ensure that no residual gases from the conducting adhesive resin (Eccobond 56), which was used for assembling the ceramic chip components and electrical contacts, contaminated the substrate. Once cooled down to room temperature, the same heating procedure was repeated for the substrate mounted on the carrier chip. After these steps nearly no organic contaminations in the SEM chamber, as inferred from mass spectrometry, were detected. The remaining dominating contaminant was then water. Therefore, we employed a home-made liquid-nitrogen trap filled with zeolite powder for $3$~hours. The decrease of the water partial pressure was monitored by mass spectrometry. Mass spectrograms after different stages of the cleaning procedure are shown in Fig.~\ref{fMS}.

In a last step, we irradiated two $1 \times10~\mu$m$^2$ areas on the substrate at $5$~kV beam voltage and $0.5$~nA beam current, at $24^\circ$C and $250^\circ$C, respectively, to check for spurious carbon deposition. In parallel mass spectrograms were recorded showing no increase in C-based impurity levels. After cooling down to room temperature the resulting vacuum base pressure was $3.55\times10^{-6}$~mbar and we considered the setup to be ready for Co-FEBID.
\subsection{Sample fabrication and post-processing}
In the Co-FEBID process the beam parameters were $5$~kV/$0.5$~nA, the pitch was $20$~nm, the dwell time was $50~\mu$s, the precursor temperature was $27^\circ$C, and the process pressure was $8.16 \times 10^{-6}$~mbar for a needle position of the gas injector at $70~\mu$m height and $70~\mu$m lateral shift from the writing field position. In order to avoid undesired spontaneous Co-growth~\cite{Mut12bjn} we kept the working pressure below the $10^{-5}$~mbar range.

\begin{figure}
\centering
   \includegraphics[width=0.7\textwidth]{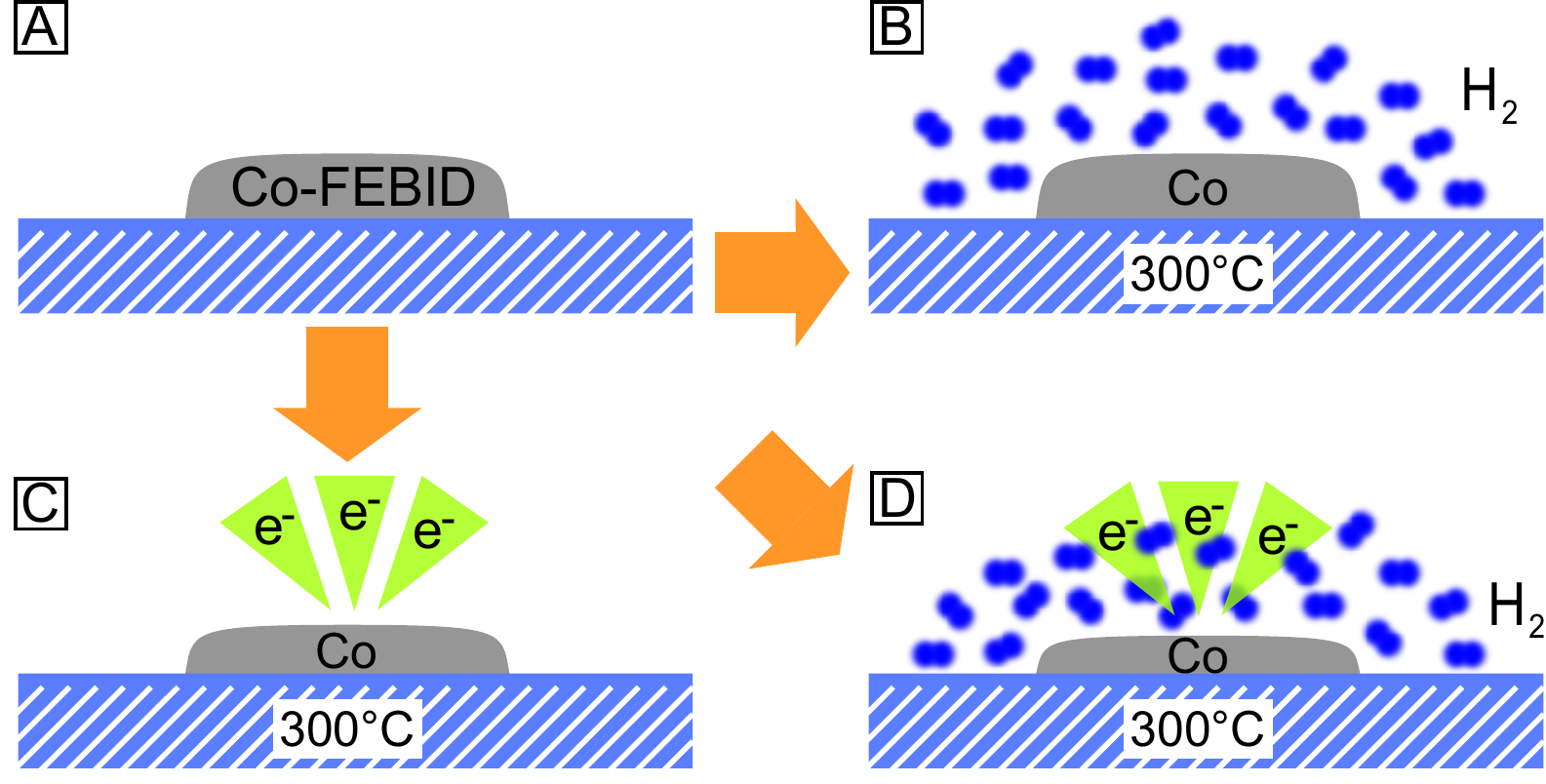}
    \caption[]
    {Schematics of the preparation of the samples used in this work. Sample A was left as-deposited and was not heated, whereas samples B, C, and D were. The heating was accompanied by a H$_2$ flux let into the SEM chamber for sample B, by electron irradiation for sample C, and by a combination of both these for sample D. The process parameters are detailed in the text.}
   \label{fChart}
\end{figure}

The samples are four strip-shaped Co-based deposits subject to different in-situ post-processing treatments. Throughout the text the samples will be referred to as sample A, B, C, and D, in accordance with the chart in Fig.~\ref{fChart}. Sample A was left as-deposited for reference purposes, whereas the others were heated to $300^\circ$C and underwent post-processing. Specifically, sample B was subject to a H$_2$ flux fed into the SEM chamber through a home-made gas injection needle positioned at $100~\mu$m height and $100~\mu$m lateral shift from the writing field position, up to a pressure of $1.5 \times10^{-5}$~mbar for 30 minutes. Sample C was irradiated with $5$~kV/$0.5$~nA electrons, with a dose of $100$~nC/$\mu$m$^2$, in vacuum at a pressure of $4.5 \times10^{-6}$~mbar. Sample D was irradiated with the same dose as sample C and in the presence of the same H$_2$ atmosphere as sample B. One remark is in order concerning the choice of the substrate temperature. In several preliminary studies we found that purification at lower temperatures was not effective enough and results were best for $300^\circ$C which is the upper temperature limit for our heating setup.
\begin{figure}
\centering
   \includegraphics[width=0.8\textwidth]{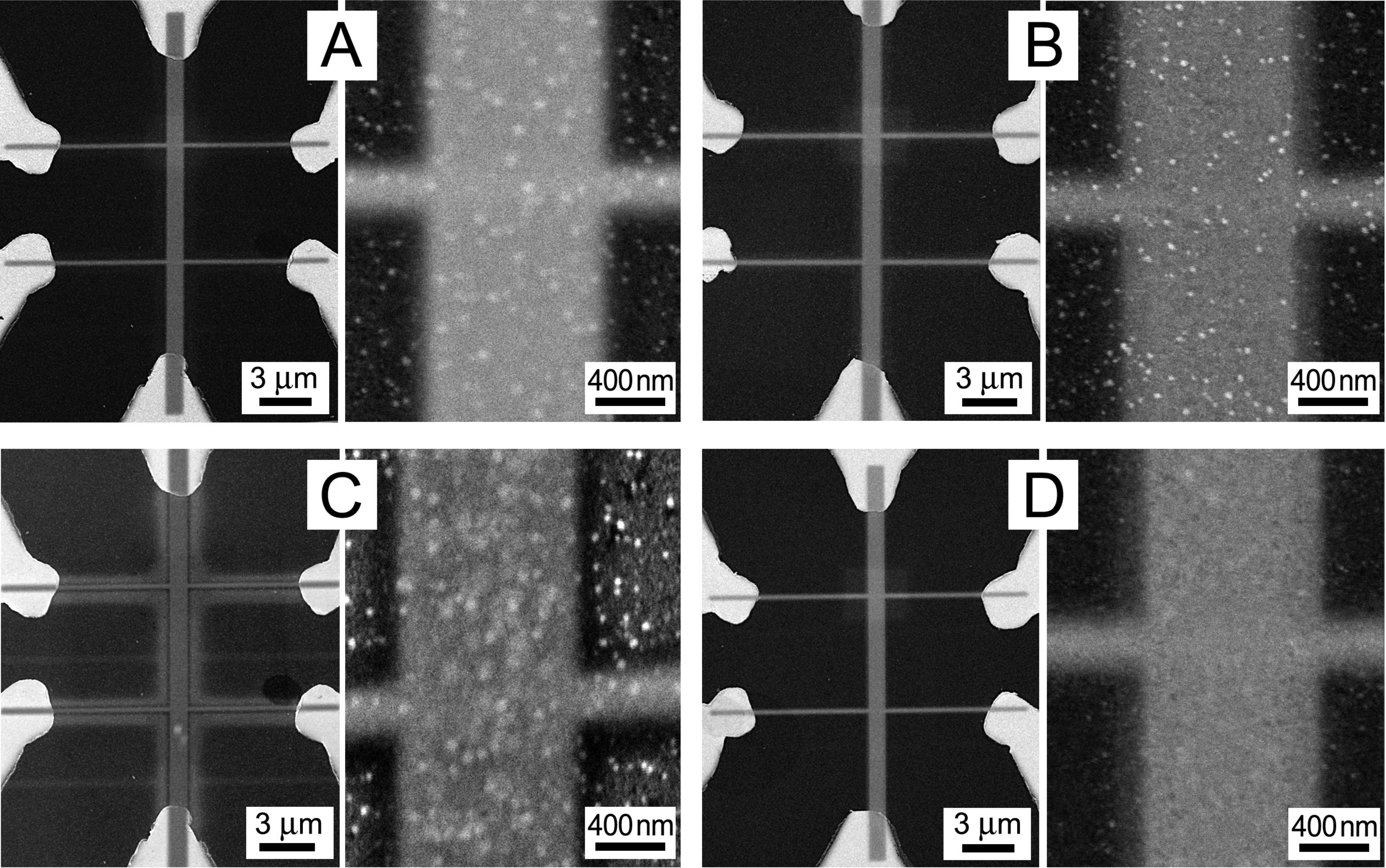}
    \caption[]
    {SEM images of the fabricated samples: Three pairs of Au contacts are bridged with Co-FEBID stripes. In electrical resistance measurements, the current is applied along the wide stripe and the two pairs of side stripes serve as voltage leads for measuring the longitudinal and transverse (magneto-)resistance components.}
   \label{fSEM}
\end{figure}

SEM images of the samples thus fabricated are shown in Fig.~\ref{fSEM}. The deposits are 6-contact stripes with an area between the voltage leads of $7~\mu$m in length and of $1~\mu$m in width, while the width of the voltage leads amounts to $100$~nm. The nominal as-deposited thickness of all samples was $50\pm2$~nm, while their thicknesses after the post-growth processing inferred from atomic force microscopy are reported in Table~\ref{tGeom}. The different roughnesses of the samples is associated with the purification treatment and will be discussed below in the context of microstructural characterization by using transmission electron microscopy (TEM). We also note that the presence of halos is not a regular result for the purification approach C and we attribute this to the temperature- and e-beam-stimulated autocatalatic growth of Co~\cite{Mut12bjn}.
\begin{table*}[t]
   \centering
   \begin{tabular}{l*{8}{c}}
    Sample   & Post-processing                      & $d$,            & $\rho_\mathrm{280K}$, & RRR    & $M_s$,  &  $\rho_{AH,s}$       & $R_{OH}$        \\
                &                                           & nm              & $\mu\Omega$cm          &                & T         &  $\mu\Omega$cm   & $\mu\Omega$cm/T   \\
\hline
    A          & as-deposited                          & 48              & 280                           & 1       & 1          &  1.85                    & 0                         \\
    B          & $300^\circ$C + H$_2$              & 51              & 62.4                          & 1.3     & 1.2       &  0.342                    & -0.005                    \\
    C          & $300^\circ$C + e$^*$             & 38              & 38.0                          & 1.4     & 1.5       &  0.17                     & -0.007                    \\
    D          & $300^\circ$C + H$_2$ + e$^*$  & 31              & 22.4                          & 1.5     & 1.7       &  0.077                   & -0.005                    \\
   \end{tabular}
   \caption{Parameters of samples. $M_s$ stands for the saturation magnetization determined at the intersection point of the anomalous $\rho_{AH}$ and ordinary $\rho_{OH}$ contributions to the Hall resistivity $\rho_H(H)$, as depicted below in Fig.~\ref{fHall}. See text for details.}
   \label{tGeom}
\end{table*}

\subsection{Material composition analysis}
The thickness-integrated material composition in the samples was inferred from energy-dispersive X-ray (EDX) spectroscopy in the same SEM, without exposure of the deposits to air. The EDX parameters were $3$~kV and $1.6$~nA. Here the beam energy, which must be not less than twice the energy for the material to be detected, determines the effective thickness of the layer being analyzed. In this experiment the probed thickness amounts to approximately $35$~nm, as calculated by the simulation program Casino~\cite{Casino}. Given the deposit thicknesses reported in Table~\ref{tGeom}, this corresponds to approximately 90\% of the electron beam energy dissipated in the deposits A, B, and C, while it is $85\%$ for sample D. The material composition was calculated taking into account ZAF (atomic number, absorbtion and fluorescence) and background corrections. The software we used to analyze the material composition in the deposits was EDAX's Genesis Spectrum v.5.11. The statistical error in the elemental composition is 1.5\%.

\begin{figure}
\centering
   \includegraphics[width=0.47\textwidth]{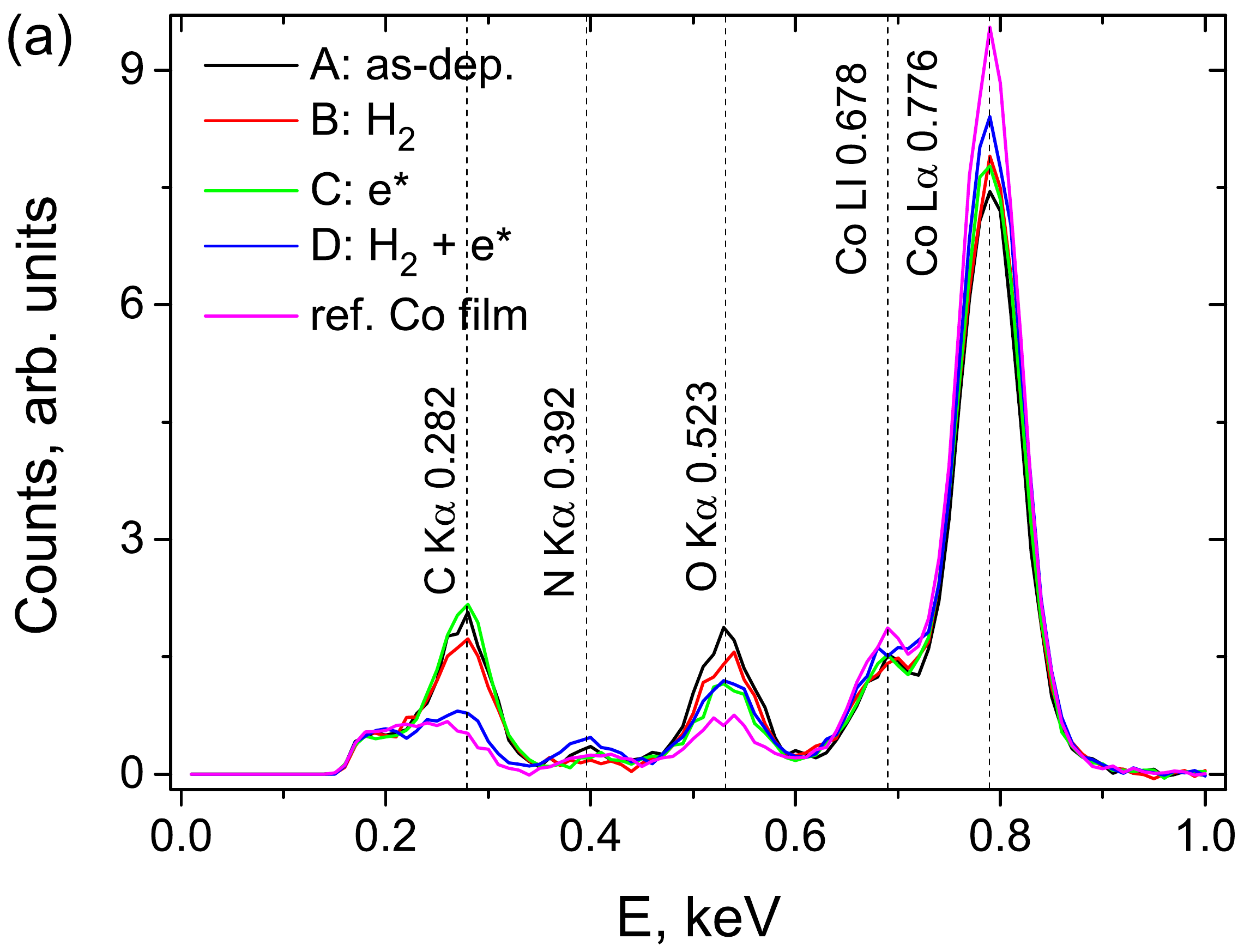}\hspace{4mm}
   \includegraphics[width=0.48\textwidth]{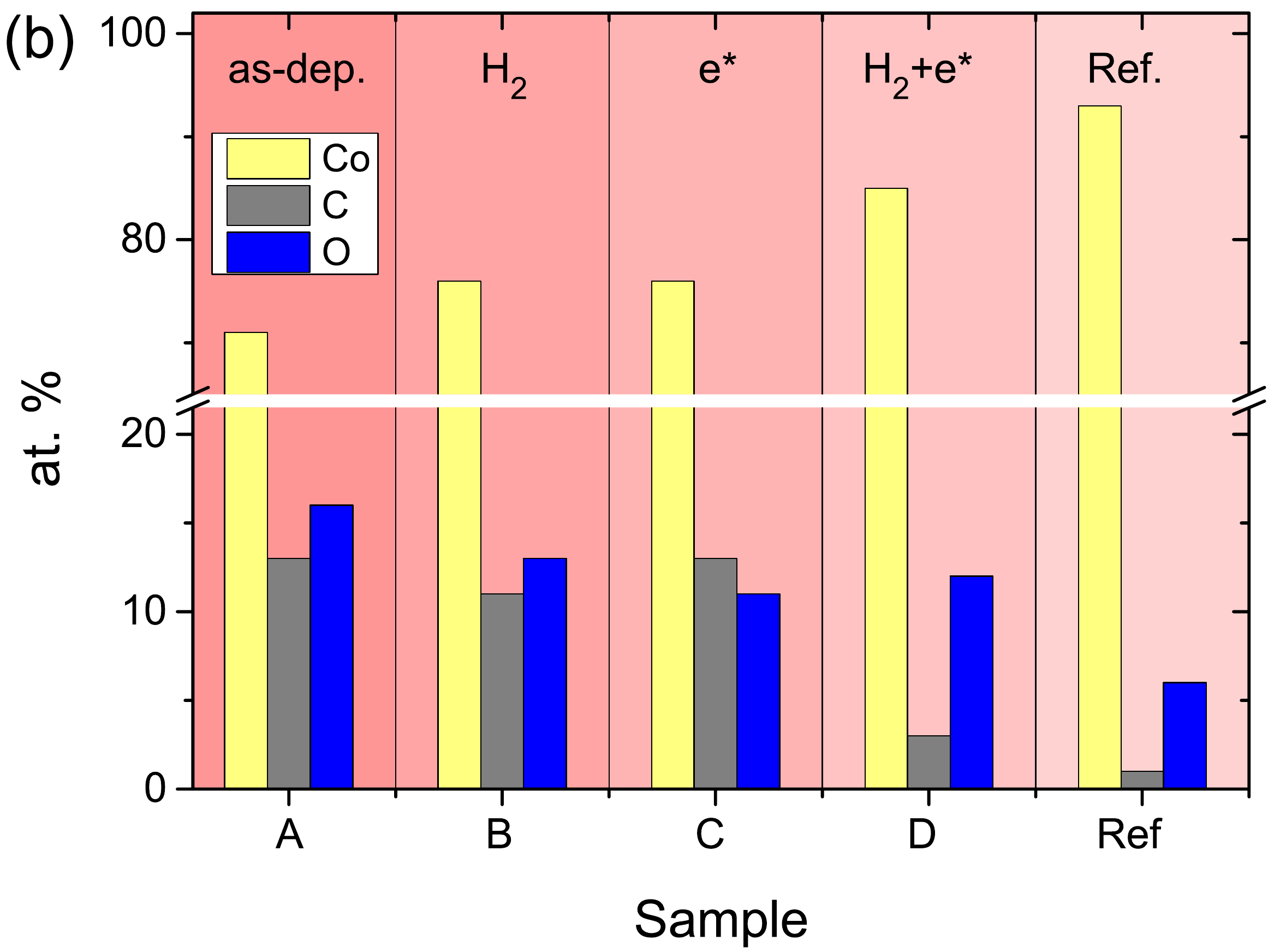}\vspace{2mm}
    \caption[]
    {Energy-dispersive X-ray spectrograms (a) and the quantified material composition (b) in all samples. In the sample caption in the right panel ``Ref.'' stands for the reference Co film grown by PVD.}
   \label{fEDX}
\end{figure}

The EDX spectra in Fig.~\ref{fEDX}(a) demonstrate peaks of four elements: Co, C, O, and N. The peak corresponding to N arises due to the relatively thin strip thickness, so that a contribution from the topmost layer of the substrate (Si$_3$N$_4$) cannot be avoided. Also, for reference purposes we have included in Fig.~\ref{fEDX}(a) the EDX spectrum of a $450$~nm-thick Co film grown on a Si/SiO$_2$/Si$_3$N$_4$ substrate by physical vapor deposition (PVD) at a base pressure of $3\times10^{-7}$~mbar and a growth rate of about $1~\mathrm{\AA}$/sec using Co of 99.99\% purity. The quantified data in Fig.~\ref{fEDX}(b) are normalized in such a way that the background has been subtracted, the N peaks have been excluded from the quantification, and all EDX spectra have been reduced to show equal areas below the curves. We now compare the data in Fig.~\ref{fEDX}(b) in detail.

The reference test made on the epitaxial Co film shows 93 at.\% of Co, 6~at.\% of O, and  1~at.\% of C. The 6~at.\% of O is attributed to an oxide layer formed on the film surface. The test made on the as-deposited sample A shows 71 at.\% of Co which is the poorest metal content among all probed samples. Samples B and C exhibit an increase of the Co content by about $5$~at.\% with respect to the as-deposited sample A. This is at the expense of the contents of C and O in sample B, while the irradiated sample C shows the same C content of 13~at.\% as the as-deposited sample A. Finally, sample D, which underwent a combination of purification treatments, exhibits a Co content of $85$~at.\%, $12$~at.\% of O, and an only minor contribution of C of $3$~at.\%. In all, the implementation of the in-situ purification treatments points to the continuous increase of the metal content in the samples and the effective removal of carbonaceous ligands. The latter conclusion is supported by a decrease of the thickness of the processed samples which has indeed been confirmed by atomic force microscopy. The most pronounced reduction of the thickness by about $30$~\% with respect to the as-deposited sample A has been observed for sample D.

\subsection{Microstructural characterization}
In order to compare the microstructure of the as-grown and purified samples transmission electron microscopy (TEM) measurements were done for an additional series of 80\,nm-thick Co-FEBID samples fabricated with the same FEBID parameters as samples A-D and purified according to the respective protocols A-D in Fig.~\ref{fChart}. The four samples for the TEM measurements, denoted by A$^\prime$-D$^\prime$ in what follows, were covered with a 300\,nm-thick protective Pt-C layer deposited on top of all samples by FEBID.

TEM and STEM images were taken in a TEM (FEI, Tecnai F20) with a Schottky gun operating at 200 kV. Electron energy loss spectroscopy (EELS) data were obtained with a post-column energy filter from Gatan (GATAN, USA) and a 2k CCD. For the EELS measurements the STEM mode was used for an exact position and correlation of the electron beam with the sample and its composition. An energy dispersion of 0.2\,eV/channel was chosen to see the carbon K-edge and the oxygen K-edge in the same EEL-spectrum. To extract the signals for carbon and oxygen a power law model was used for subtracting the background. The background window size was 20 eV for carbon and 50 eV for oxygen, while the signal window size was 9 eV and 25 eV, respectively.
\begin{figure}[t!]
\centering
   \includegraphics[width=0.95\textwidth]{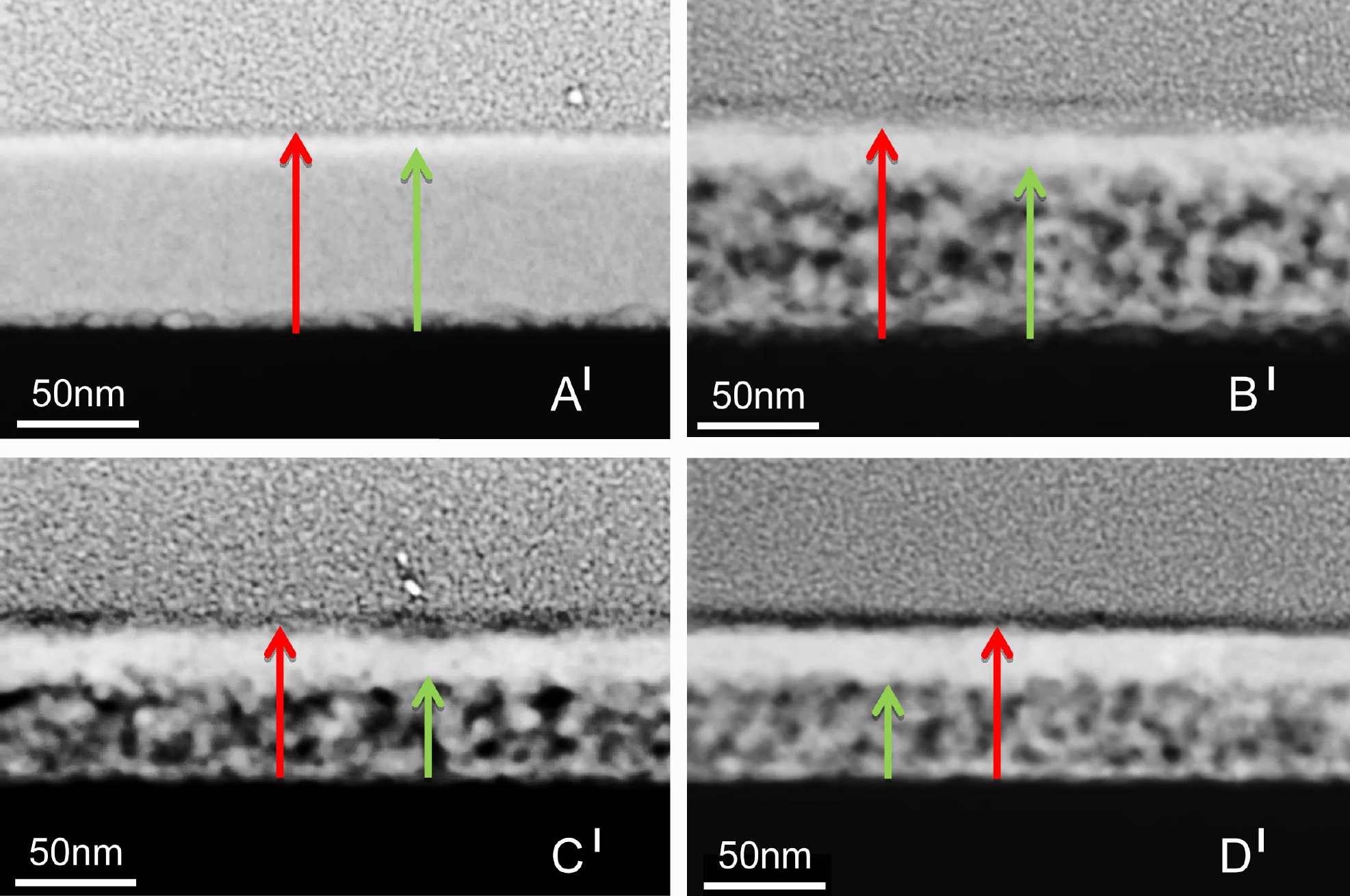}
    \caption[]
    {TEM micrographs of the four samples acquired in the high angle annular dark field mode.}
   \label{fADF}
\end{figure}

\begin{figure}
\centering
   \includegraphics[width=0.48\textwidth]{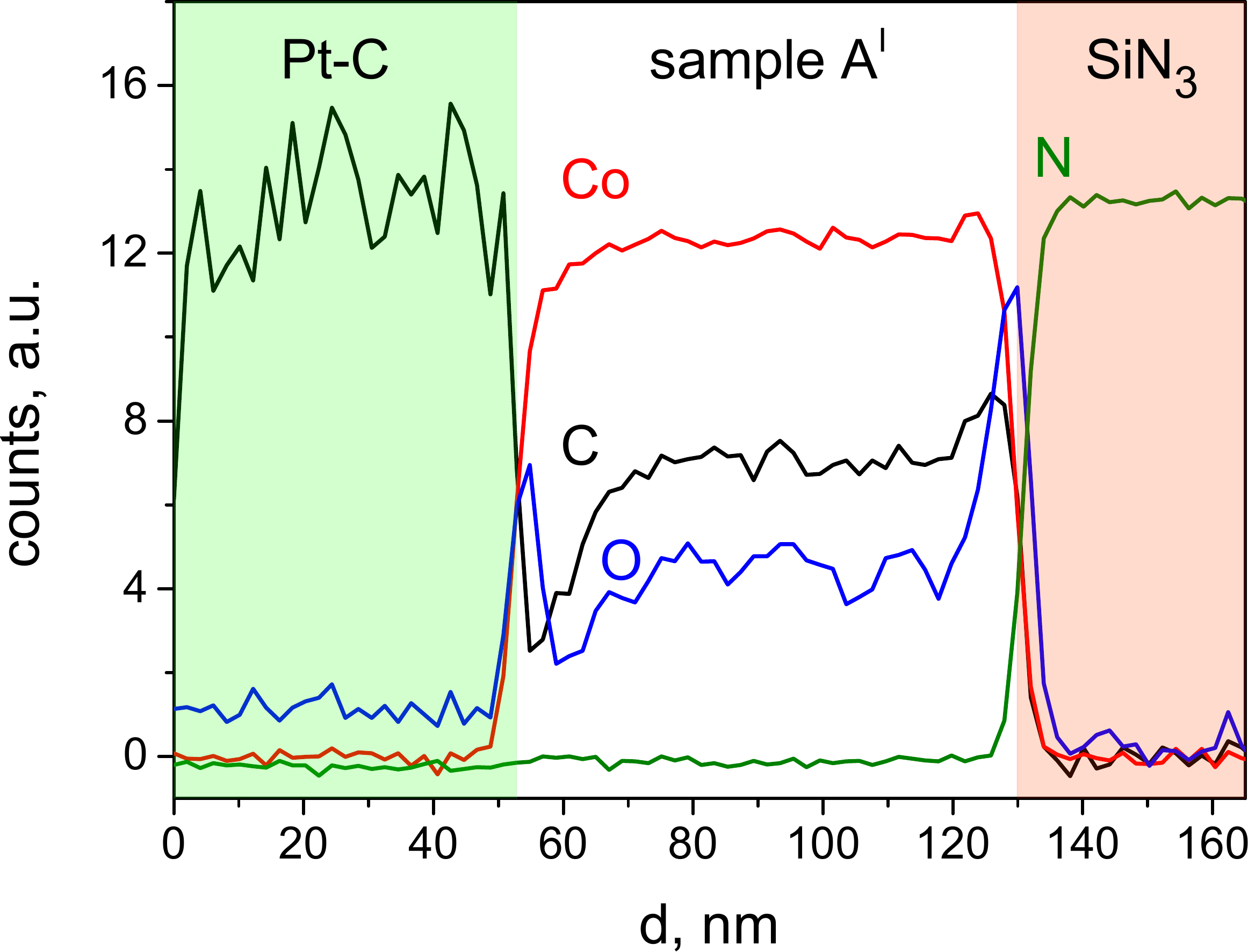}
   \vspace{0.3cm}\hspace{0.3cm}
   \includegraphics[width=0.48\textwidth]{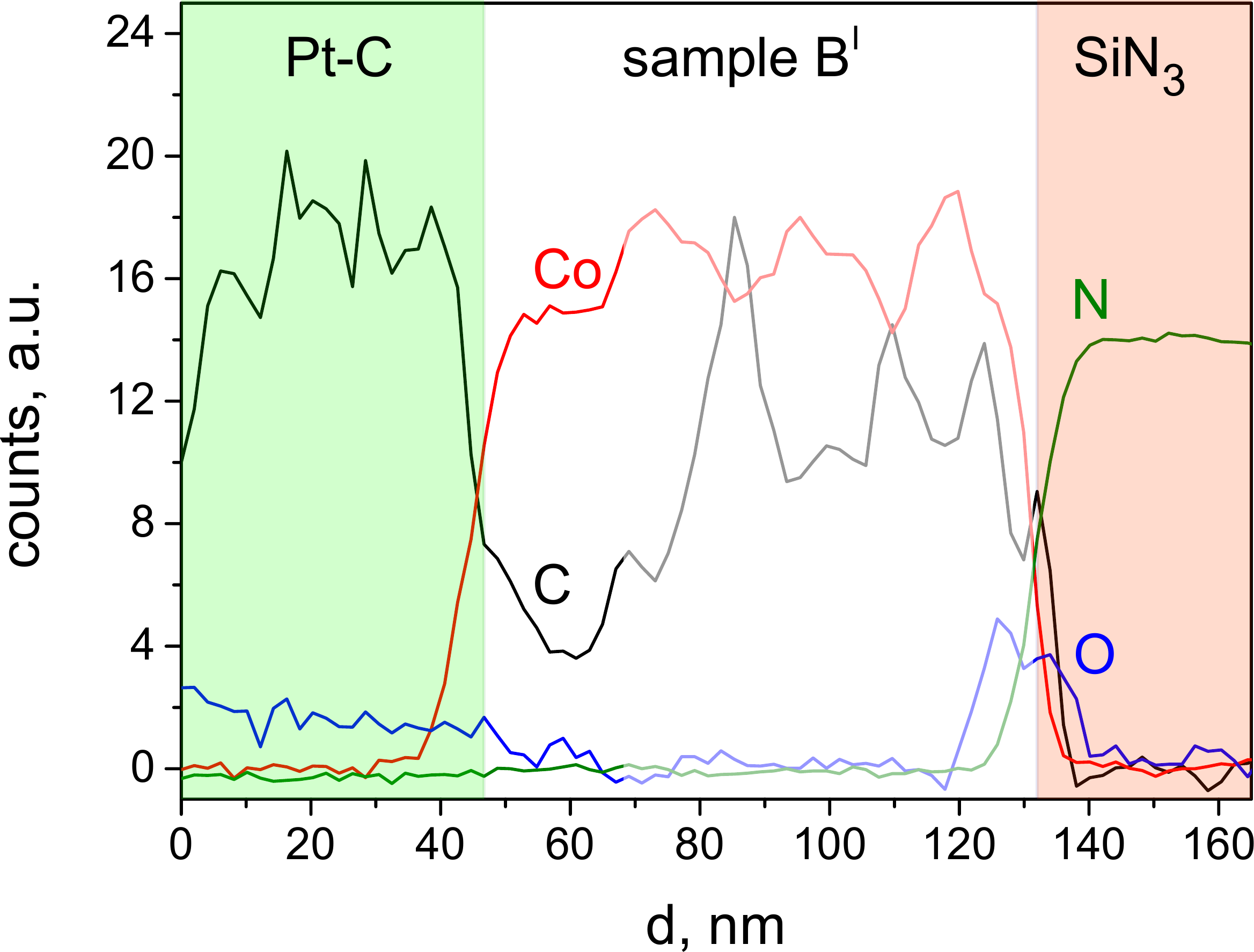}
   \includegraphics[width=0.48\textwidth]{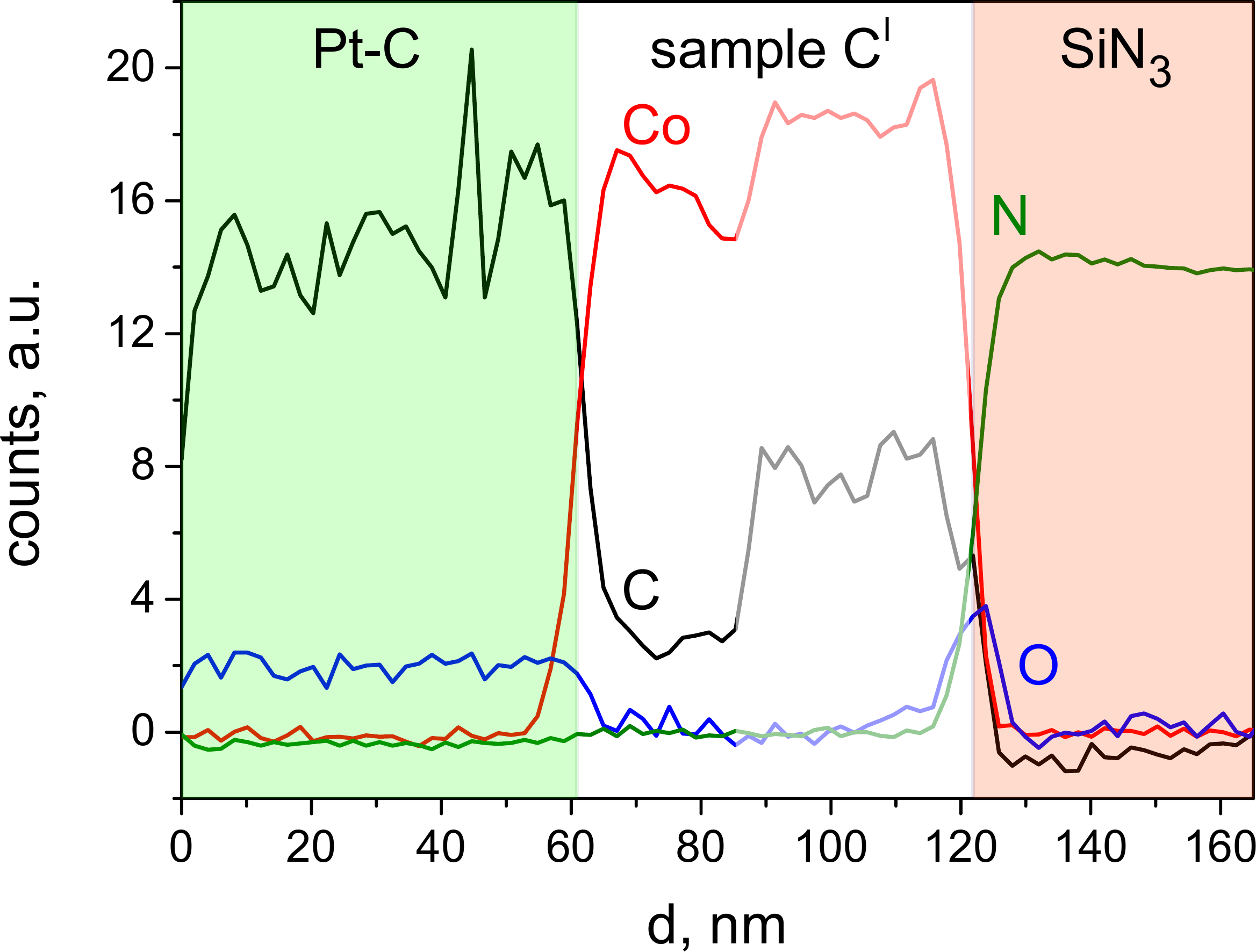}
   \hspace{0.3cm}
   \includegraphics[width=0.48\textwidth]{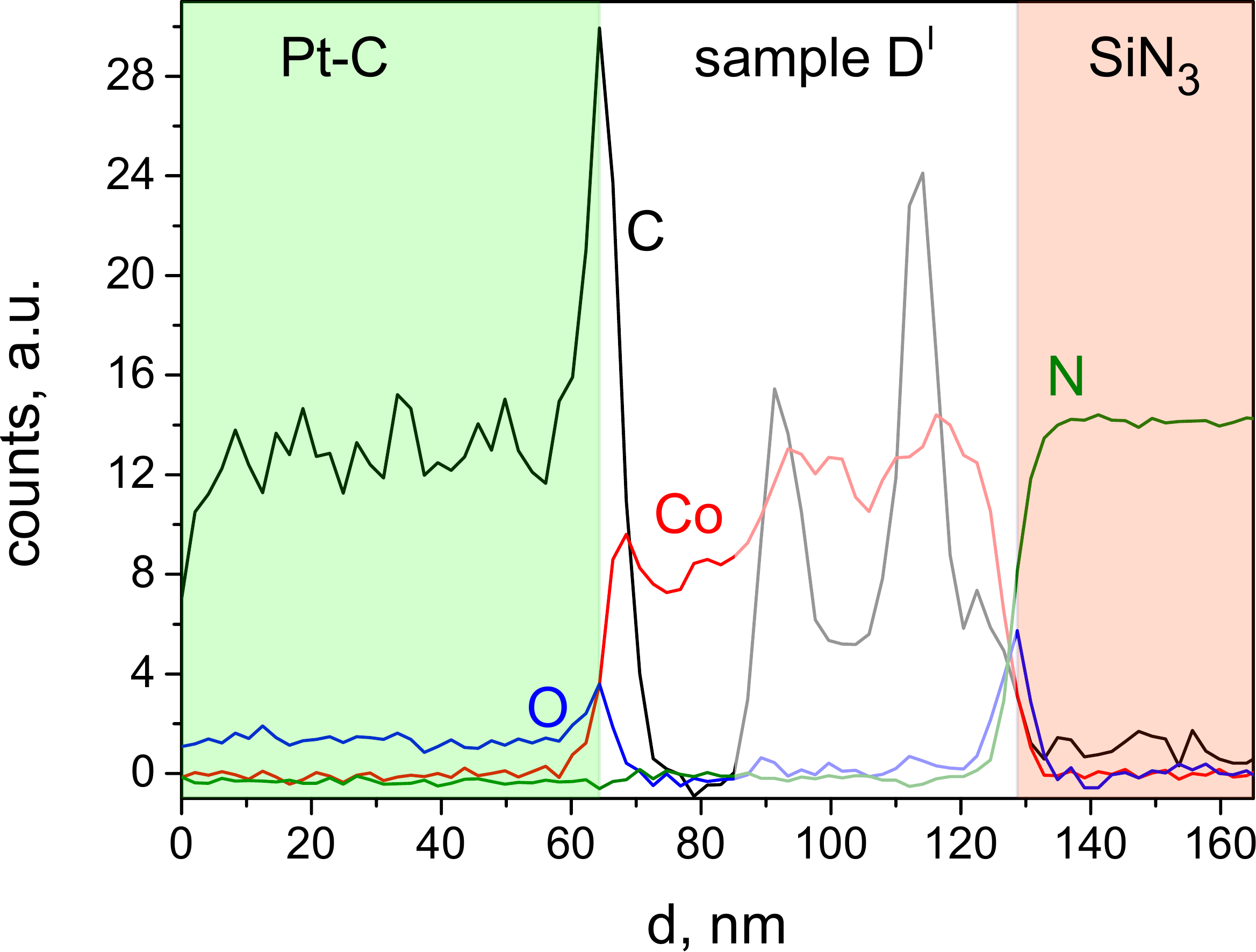}
    \caption[]
    {EELS-STEM spectrograms for the four samples. Pt-C denotes the protective layer on top of the samples, while SiN$_3$ the top layer of the substrate. For samples B$^\prime$, C$^\prime$ and D$^\prime$ the spectra referring to the porous regions are in damped colors because of their reduced reliability. See text for details.}
   \label{fELS}
\end{figure}

Figure~\ref{fADF} presents cross-sectional TEM images of all samples in the high angle annular dark field mode. The respective spectrograms obtained by STEM-EELS are shown in Fig.~\ref{fELS}. We now consider the TEM and EELS data in detail. From the micrograph in Fig.~\ref{fADF}(a) it follows that sample A$^\prime$ is homogenous over its entire thickness, showing just a slightly enhanced content of oxygen within an about 5\,nm-thick topmost layer, see Fig.~\ref{fELS}(a). The sample is continuous and has a flat, fine-grained morphology. By contrast, in Figs.~\ref{fADF}(b-d) in samples B$^\prime$, C$^\prime$, and D$^\prime$ one recognizes two layers. In sample B$^\prime$ the about 10\,nm-thick top layer is a highly compact Co layer with some residual carbon, while the thicker bottom layer exhibits a pronounced porous structure. In Fig.~\ref{fADF}(c) the irradiated sample C$^\prime$ is also porous, but in contrast to sample B$^\prime$, its Co-rich top layer is thicker (about 17\,nm) and the degree of porosity (the number and sizes of pores) of the bottom layer is smaller. Sample D$^\prime$ shows the thickest top layer with rather well-defined interfaces. This layer appears as C- and O-free Co with a thickness of 20\,nm. The bottom layer is still porous, but this porosity is much less pronounced than in samples B$^\prime$ and C$^\prime$. This porosity leads to complications in interpreting the STEM-EELS data further, as the sampled volume and the associated count rates depend on the actual TEM lamella thickness at the respective probing position. What can be stated for samples B$^\prime$-D$^\prime$ is that the overall O content is clearly reduced. The compact top-layer of sample D$^\prime$ appears to O-free within the spectral accuracy. Also, despite of the strong variations in the C and, to a lesser degree, Co count rate in the porous layer, the absence of a C signal in the compact top layer of D$^\prime$ is apparent.

From these observations we identify protocol D in Fig.~\ref{fChart} to be the most effective with regard to purifying Co-FEBID structures. Also, electron irradiation leads to compacter structures, as is evident for samples C$^\prime$ and D$^\prime$, which show both a height reduction amounting to about 25\,$\%$ for sample D$^\prime$.
\section{Electrical transport measurements}
\subsection{Resistivity}

Transport measurements were carried out in a helium-flow cryostat equipped with a superconducting solenoid. The electrical resistance was measured as a function of temperature in the standard 4-probe geometry where all four electrodes were made from Co-FEBID. The electrical and magneto-resistance measurements were done in the dc current mode, with a current density of the order of $10$~kA/cm$^2$. The dc current was supplied by a Keithley~2636A source-meter and the dc voltage was measured with an Agilent~34420A nanovoltmeter.

\begin{figure}
\centering
   \includegraphics[width=0.58\textwidth]{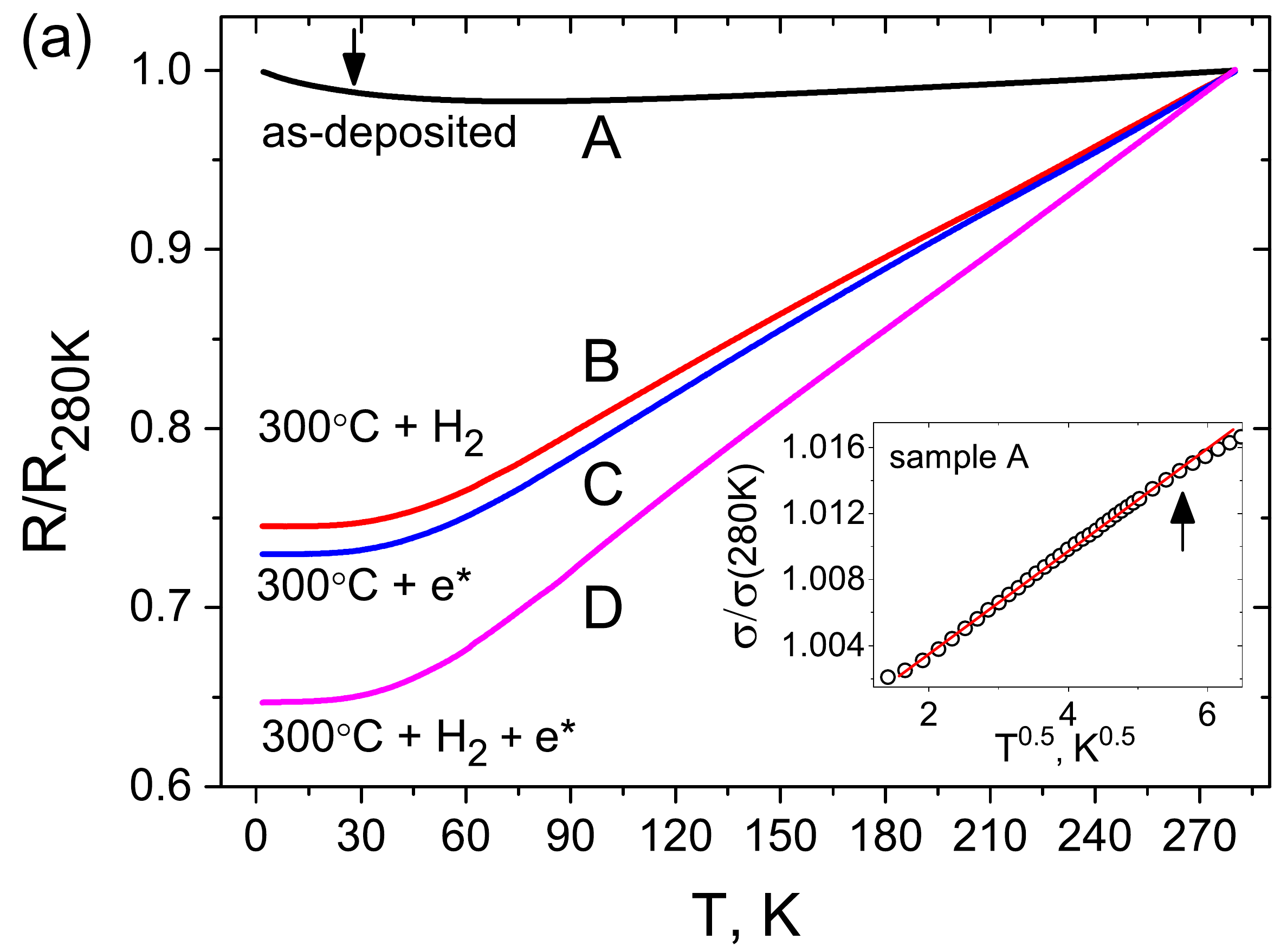}\\[1cm]
   \includegraphics[width=0.60\textwidth]{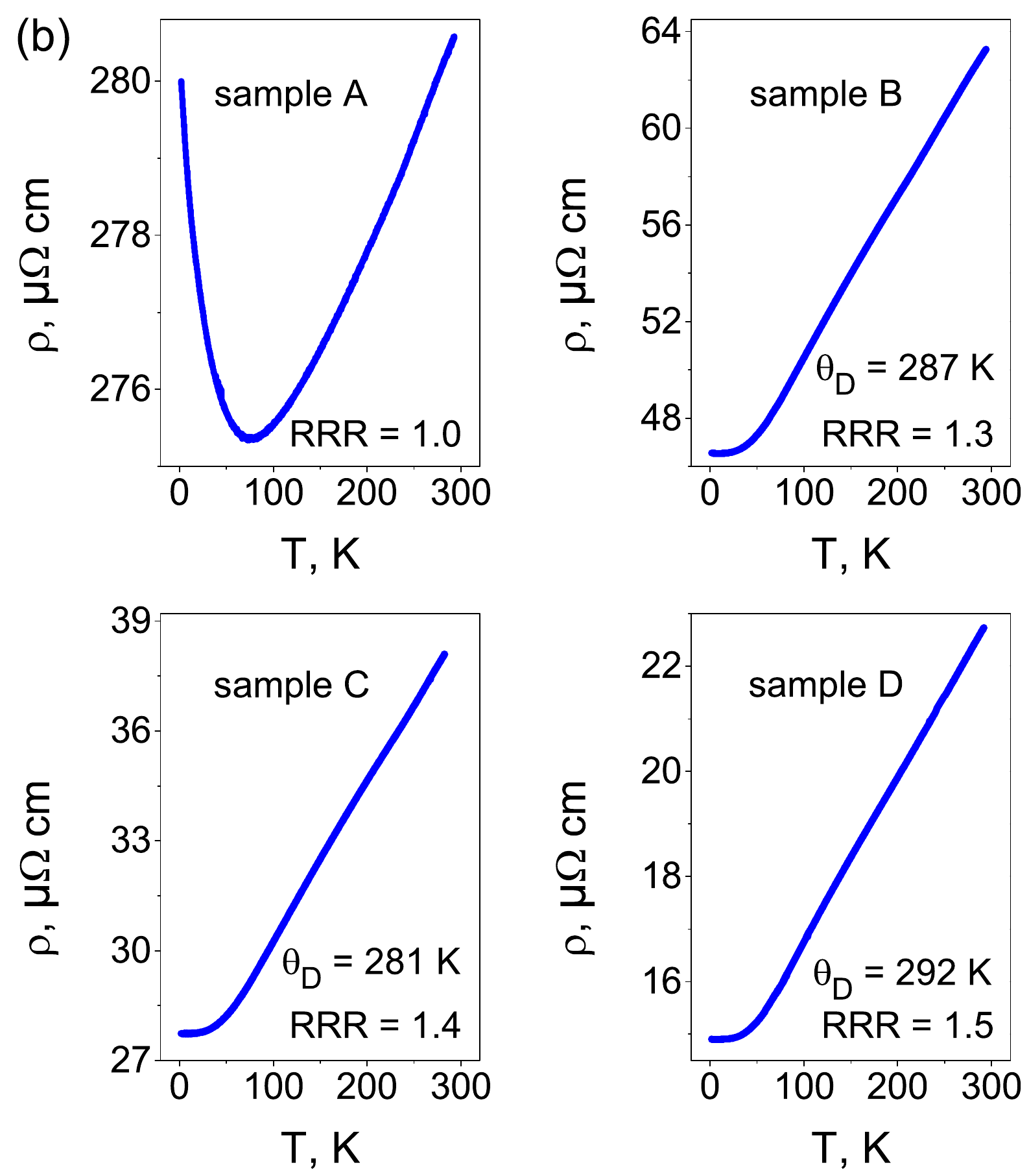}
    \caption[]
    {(a) Cooling curves for all samples, as indicated. Inset: Low-temperature conductivity data in $\sigma$ vs $T^{0.5}$ representation. The solid line is a fit to the law $\sigma \propto T^{0.5}$, while the arrows mark the temperature where the data start to deviate from this law. (b) Resistivity for all samples, as indicated.}
   \label{fRvT}
\end{figure}

The temperature dependences of the electrical resistance of samples A--D are shown in Fig.~\ref{fRvT}(a). The as-deposited sample A shows a very weak metallic behavior that gradually develops a localization-induced increase below about $70$~K. We note that its resistivity $\rho_{\mathrm{280K}} = 280~\mu\Omega$cm is just slightly smaller than the resistivity of deposits formed in the result of spontaneous dissociation of the same precursor~\cite{Mut12bjn}. We attribute this high resistivity to the granular microstructure of the sample as follows. In the inset of Fig.~\ref{fRvT} the square-root temperature dependence of the normalized conductivity is presented for sample A. The curve shows a linear behavior in the temperature range from about $3$~K up to $28$~K. This behavior is in agreement with the transport theory for ordered granular metals in the strong intergrain coupling regime proposed by Beloborodov $et~al.$~\cite{Bel03prl,Bel07rmp}. We note that similar temperature-dependent data were also reported for nanogranular Pt-C samples prepared by FEBID, where $\sigma\sim T^{0.5}$ was observed in the same temperature range~\cite{Sac11prl}. In the following, and in particular with a view to interpreting the magnetoresistance and Hall effect data, we consider sample A to be a granular metal in the strong-coupling limit.

By contrast, all purified samples exhibit improved electrical conducting properties, as their integral resistance drops by a factor of $1.3$-$1.5$ while cooling down, in agreement with results for polycrystalline Co films~\cite{Kot05prb}. The resistivity values reported in Fig.~\ref{fRvT}(b) and Table~\ref{tGeom} were calculated assuming a homogenous material throughout the complete layer thickness. The resistivity of the processed sample D is by more than one order of magnitude smaller than that of the as-deposited sample A and is a factor of $4$ larger than the reference bulk value of $5.8\,\mu\Omega$cm for Co~\cite{Kit04boo}. If we consider the results of the TEM investigations on the reference sample D$^\prime$, we have to assume a top layer of pure Co metal of about 20\,nm thickness on top of a 11\,nm thick residual layer with porous microstructure which we expect to have a higher resistivity. For such a bilayer structure we can provide an upper bound of the resistivity for the Co layer of $14.5\,\mu\Omega$cm ignoring conductance contributions from the bottom layer and expected resistance contributions from the interface region.

Despite of the layered structure of samples B--D their $\rho(T)$ curves can be rather well fitted to the Bloch-Gr$\ddot{\mathrm{u}}$neisen formula~\cite{Zim60boo}
\begin{equation}
    \label{BG}
    \rho(T) = \rho_0 + K(T/\theta_D)^n \int_{0}^{\theta_D/T}dx\frac{x^n}{(e^x-1)(1-e^{-x})},
\end{equation}
where $\theta_D$ is the Debye temperature, K is a fitting parameter, and $n$ is an integer determining the power law which in turn depends on the prevailing scattering mechanism in the sample. The fitting parameter $K$ is chosen such that the best possible coincidence with the experimental curves in Fig.~\ref{fRvT} is achieved for $\rho_{10\mathrm{K}}$ and $\rho_{280\mathrm{K}}$. All curves $\rho(T)$ are fitted by equation~(\ref{BG}) with $n=5$ which implies that the resistance is due to electron-phonon scattering~\cite{Zim60boo}, while electron-magnon contributions $\propto T^2$ are small, which is in-line with measurements on polycrystalline Co thin films~\cite{Kot05prb}. Varying the Debye temperature as a fitting parameter, the best possible coincidence of the measured data to expression~(\ref{BG}) is achieved with Debye temperatures between $280-295$~K. The Debye temperature for bulk Co is $385$\,K~\cite{Ash06boo}. We attribute the reduced values deduced from our analysis to the layered structure of the samples in conjunction with finite-size effects, as a similar reduction of $\theta_D$ has been observed for thin Pt layers~\cite{Sac14ami}.
\subsection{Magnetoresistance}
\begin{figure}
    \centering
    \includegraphics[width=0.5\textwidth]{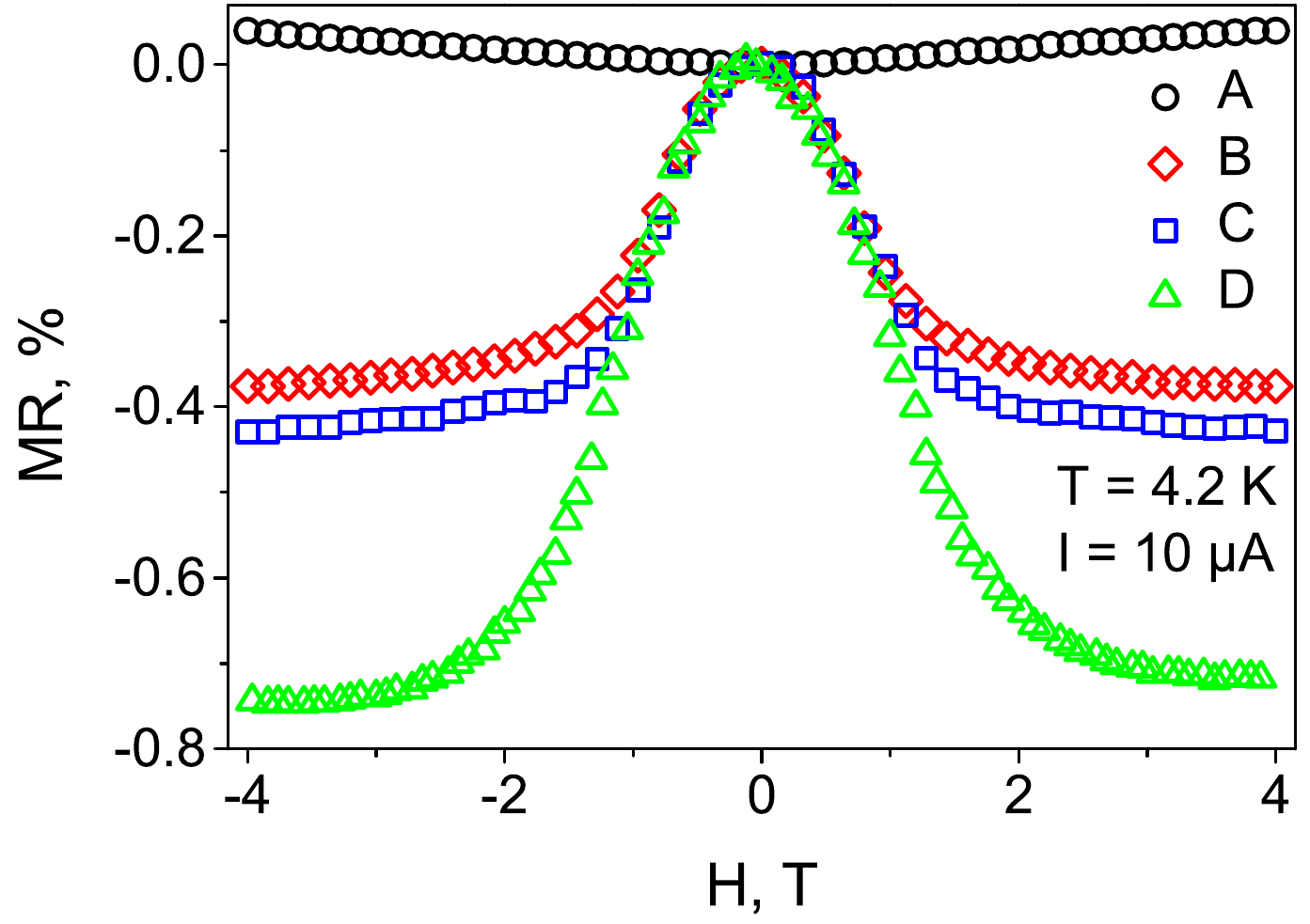}
    \caption[]
    {Perpendicular magnetoresistance measurements at $4.2$~K for all samples. MR is defined as $100[R(H) - R(H = 0)]/R(H = 0)$.}
    \label{MR}
\end{figure}

Magnetoresistance (MR) and Hall effect measurements were done at $4.2$~K with magnetic fields up to $\pm4$~T applied perpendicular to the substrate plane. The results for all samples are shown in Fig.~\ref{MR}. MR is defined as $100[R(H) - R(H = 0)]/R(H = 0)$, where $R(H)$ and $R(H = 0)$ are the resistances at a given magnetic field $H$ and zero field, respectively.

We observe a positive MR for sample A, whereas the MR is negative for samples B--D. Due to the layered structure of the purified samples, the interpretation of the magnetoresistance results has to remain on the qualitative level.

A positive magnetoresistance for a granular metal, as observed for sample A, can be attributed to the influence of the magnetic field on the wave function attenuation length for the electronic surface states of the Co grains that are subject to tunnel-coupling to neighboring grains. For granular Pt-FEBID structures it was shown that the wave-function shrinkage model, that predicts a reduction of the attenuation length with increasing magnetic field, can account for the observed positive MR~\cite{Por14pcm}.

For samples B--D, all showing a negative MR, some distinctive features are discernible. Firstly, the value of the MR at $4$~T of sample D is about a factor of two larger than that of samples B and C. Secondly, the MR of sample D already saturates at $\pm3$~T, whereas for samples B and C no such saturation is observed. This MR behavior of samples B and C is reminiscent of that observed in materials with competing anisotropic MR (AMR) and intergranular tunneling MR (ITMR), e.g. iron microwires grown by FEBID~\cite{Cor12jpd}. The AMR originates from an anisotropic electron scattering due to spin-orbit coupling inside metallic Co, while the ITMR is caused by spin-polarized electrons tunneling between metallic grains embedded in an insulating matrix. The ITMR is often difficult to saturate even under very high magnetic fields~\cite{Cor12jpd}, while the AMR is saturating due to ordering of the magnetic domains. From our TEM studies, that show for samples B$^\prime$ and C$^\prime$ still a substantial amounts of C in the Co top layer, a residual granularity of this layer appears to be likely, which would lend support to the competing AMR and ITMR interpretation. In contradistinction to samples B and C, sample D shows a pure AMR signal which attests to its metallic character. The AMR value of $0.74\%$ at $4.2$~K is nearly the same as in pure Co nanowires~\cite{Lev05prb}. The observed saturating behavior with a quadratic dependence at $H \lesssim H_s$ is expected for ferromagnetic materials since the electrons scattering probability decreases as more magnetic moments are aligned along the direction of the external magnetic field. These conclusions are valid under the assumptions that for sample D current transport is mostly limited to the top layer of 20\,nm thickness which consists of pure metallic Co according to the TEM results.
\subsection{Hall effect}
\begin{figure}
\centering
   \includegraphics[width=0.75\textwidth]{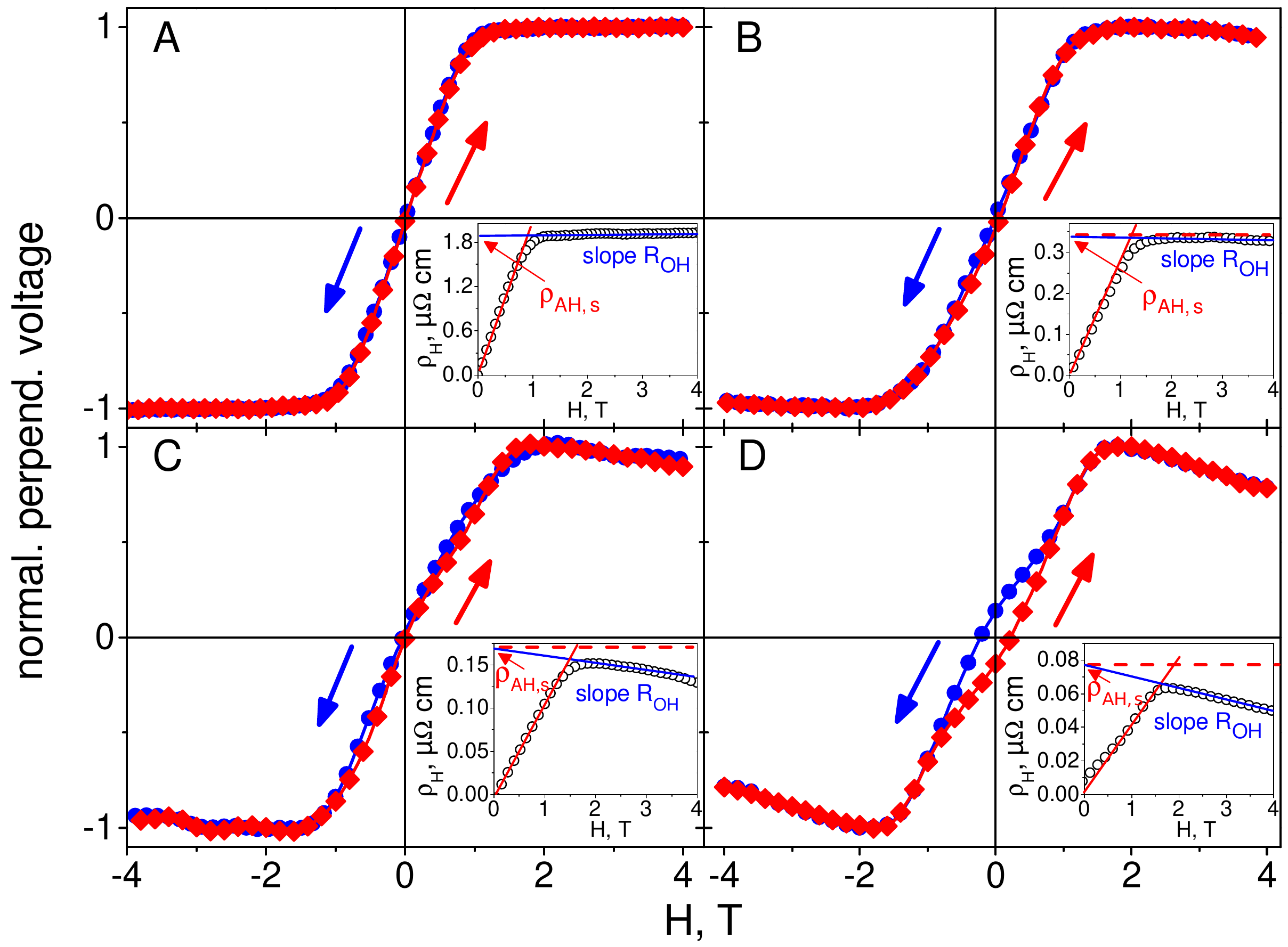}
    \caption[]
    {Isothermal Hall voltage cycling for all samples at $4.2$~K. Before measurements, all samples were magnetized in a field of $4$~T. The determination of the anomalous Hall resistivity $\rho_{AH}$ and the  ordinary Hall resistivity at $H = M_s$, $\rho_{OH}$, is indicated.}
   \label{fHall}
\end{figure}

The results of Hall effect measurements done at $4.2$~K for all samples are reported in Fig.~\ref{fHall}. The Hall voltage is determined by the standard expression
\begin{equation}
     \tilde{U}_{Hall}= \frac{U_{Hall}(H)-U_{Hall}(-H)}{2}
\end{equation}
under magnetic field reversal. The Hall resistivity is given by
\begin{equation}
    \rho_H = \tilde U_{Hall} \frac{d}{I},
\end{equation}
where $d$ stands for the thickness of the deposit, and $I$ is the current passing through the sample.

As is customary for ferromagnetic materials, we write the Hall resistivity as a sum of the ordinary, Lorentz force induced, contribution (OHE) and the anomalous contribution (AHE) which is proportional to the sample magnetization. By doing this we assume a demagnetization factor $N_z\approx 1$, as is appropriate for our sample geometry \cite{Hub08boo}
\begin{equation}
    \rho_H = \rho_{OH} + \rho_{AH} = \mu_0(R_{OH}H + R_{AH}M_z) \,.
\end{equation}
$H$ is the applied magnetic field, $M_z$ is the magnetization along the field direction $\mathbf{z}$, and $R_{OH}$ and $R_{AH}$ are the ordinary and the anomalous Hall coefficients, respectively. $R_{OH}$ was determined from the slope of $\rho_H$ within the linear regime above the saturation field $H_s = M_s$ (for $N_z=1)$, as indicated in Fig.~\ref{fHall}. $\rho_{AH}(H_s)=\rho_{AH,s}$ at saturation was deduced from linear extrapolation of $\rho_H(H)$ at $H>H_s$ to zero field, as also as indicated in Fig.~\ref{fHall}. The respective values for samples A--D are given in Table~\ref{tGeom}.

In contradistinction to sample A, samples B--D demonstrate ordinary and anomalous contributions with different signs, as commonly observed for pure Co. In addition, sample D shows a notable hysteresis with a coercive field $\mu_0H_c$ of $0.22$~T and a remanent magnetization $M_r/M_s$ of about $0.14$. Describing the overall shape of the Hall curves, below the technical saturation, $M_z \leq M_s(T)$, the perpendicular resistivity increases linearly with the field due to the linear variation of the magnetization, producing a large anomalous Hall effect~\cite{Kot05prb}. At higher fields $H \geq M_s$ the field dependence is negative and the ordinary Hall effect is observed. A high value of the anomalous Hall resistivity is observed for sample A, whereas its value for samples B, C, and D is one order of magnitude smaller and is comparable with the values reported for high-quality Co-FEBID nanowires~\cite{ Fer09jpd}.

How do these observations relate to the TEM-deduced microstructure? If, again, we assume that transport is dominated by the topmost, Co-rich or pure Co (sample D), layer, we expect to observe in the Hall data the behavior of a dirty, polycrystalline ferromagnet. Given the lateral size and thickness of the stripes, sample D should be in a multi-domain state and show hysteresis effects. For a more quantitative analysis one would also have to consider the internal field and exchange field contributions to the magnetic state caused by the Co-containing, porous bottom layers. This is beyond the scope of this work. However, some additional insight can be obtained from focusing on samples B--D.

\begin{figure}
\centering
   \includegraphics[width=0.75\textwidth]{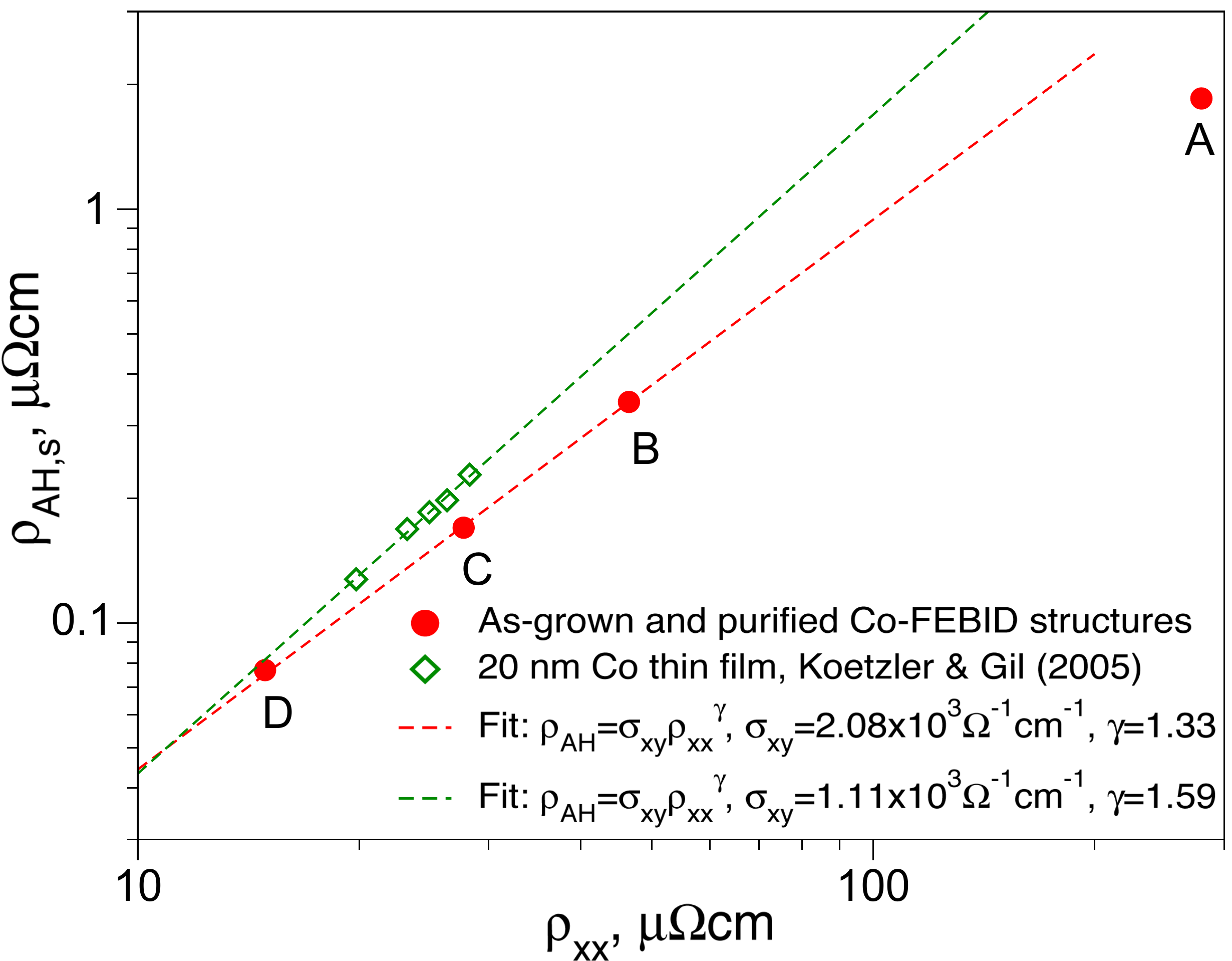}
    \caption[]
    {Anomalous Hall contribution of Hall resistivity at saturation vs.\ longitudinal resistivity at $4.2$~K for samples A--D, as indicated. For reference, data for a 20~nm polycrystalline Co thin film are also shown (data taken from \cite{Kot05prb}). See text for details concerning the scaling fits.}
   \label{fScaling}
\end{figure}

In general, the anomalous Hall effect in ferromagnetic metals has contributions with intrinsic and extrinsic origin. The intrinsic contribution arises in a perfect periodic lattice with broken time-reversal symmetry and is due to the topological properties of the Bloch states. It is thus not depending on scattering contributions and one expects for the Hall conductivity $\sigma_{xy}\propto \tau^0$, with $\tau$ being the scattering time. The extrinsic AHE is due to different asymmetric spin-orbit scattering effects which give rise to skew scattering and side jump contributions (see \cite{Nag10rmp} for a review). The extrinsic contributions are dominating in very pure metals, as $\sigma_{xy} \propto \tau^n$ with $n=1$ for skew scattering and $n=2$ for the side jump mechanism. These contributions are not relevant in the present case. The intrinsic contribution however should be relevant and is expected to follow a scaling behavior
\begin{equation}
\rho_{AH,s} = \sigma_{xy}\rho^\gamma
\end{equation}
with $\gamma=2$ in the pure limit and $\gamma\approx 1.6\dots 1.8$ for elevated resistivity values (bad metal regime)~\cite{Nag10rmp}. In Fig.~\ref{fScaling} we plot $\rho_{AH,s}$ for all samples vs.\ the longitudinal resistivity at 4.2~K in log-log representation to see whether we can verify such a scaling behavior. For direct comparison we also show data for polycrystalline Co thin films of comparable thickness (20~nm) taken from~\cite{Kot05prb}. With all caution necessary due to the layered structure of the purified samples, we would conclude that samples B--D follow the expected scaling behavior quite well and are quite comparable to the polycrystalline Co thin films with an exponent somewhat below 2 and a Hall conductivity of very similar magnitude. We note that Co-FEBID structures with metal content above 90~at$\%$ have been reported to follow a $\gamma=2$ scaling~\cite{Fer09jpd}.

In this scaling analysis we have omitted sample A from the fit. The Hall resistivity for sample A falls clearly below the extrapolated value at such high-resistivity values. We consider it likely but cannot prove that this is due to the granular nature of sample A which has also be apparent in the temperature-dependent conductivity. Meier et al.\ have performed theoretical studies on interaction and weak-localization corrections to the AHE in granular metals with ferromagnetic grains in the strong tunnel-coupling regime~\cite{Mei09prb}. They predict a loss of the scaling behavior and in fact expect $\rho_{AH,s}$ to be independent of $\rho_{xx}$.
\section{Discussion}\label{sDiscussion}
With regard to the microstructural and chemical properties of the different purification protocols we can state the following major effects:
\begin{description}
  \item [(i)] The deposits' height is reduced which is most strongly pronounced for the protocols including electron irradiation.
  \item [(ii)] The overall Co content increases and there is a clear tendency for reducing the O and C content in the topmost layer which can be as thick as 20\,nm.
  \item [(iii)] The electrical conductivity increases in all cases and shows metal-like behavior.
  \item [(iv)] There is a change of the magnetic properties, as visible by transport measurements, that indicates a ferromagnetic behavior of the topmost layer.
\end{description}
In the following we make an attempt to put forward mechanistic explanations of the physico-chemical processes which take place in the course of the different purification treatments.

We first turn to the microstructural transformations in the purified sample C in comparison to the as-deposited sample A. We recall that sample A is a nanogranular material made of metallic grains embedded in a dielectric carbon matrix. The transport mechanism in sample A is mediated by electron tunneling between grains through the potential barrier induced by the dielectric matrix. Accordingly, the transport properties of the sample can be tuned by varying the size of the grains in the matrix and by changing the properties of the matrix itself, as it has been reported for electron irradiated Pt-FEBID structures~\cite{Sac11prl,Por11jap}. This effect was explained~\cite{Sac11prl,Por11jap} by microstructural changes associated with (i) size increase of the metallic nanocrystallites with subsequent coalescence and (ii) a transformation of the amorphous carbon in as-deposited structures into a dielectric matrix with more graphite-like near-range order and thus increased transmission in the tunneling processes in the treated structures. Similar effects may take place in sample C. Our reasoning is confirmed by both, the reduction of the entire sample volume by $20$\% and the increase of the conductivity value by a factor of $7$ with respect to the as-deposited reference sample A. In addition to this, in Ref.~\cite{Gab10nan} it was reported that by adjusting the nanoparticles size and the distance between them it is possible to tune the magnetic properties of Co-C deposits. This independent observation~\cite{Gab10nan} is also in-line with our MR measurements.

We proceed now to the non-irradiated sample B. We believe that the microstructural changes in this sample are invoked by the presence of the H$_2$ atmosphere in conjunction with the high temperature. Indeed, in Ref.~\cite{Wei98itm} an improvement was reported of the conducting properties of Co$_x$C$_{1-x}$ thin films from the insulating regime for as-grown films to the metallic regime for films annealed at a temperature of $600^{\circ}$C. Therefore, we attribute the observed improvements of the conducting properties of sample B to the thermally-invoked coarsening of the Co nanogranules. These granules tend to form a percolated network which is denser close to the surface and remains incomplete in deeper layers, as is apparent by the observed porosity in TEM. We recall that the annealing process takes place in the presence of hydrogen, whereby the purification processes may be driven by Fischer-Tropsh-like reaction~\cite{And84boo}. In this catalytic reaction, hydrogen acts as a reducing agent and the reaction products are hydrocarbons and water which will be effectively oxidizing the carbon. Hence, we believe that the reduction of the carbon and oxygen contents is caused by the formation of volatile CO. At this, the effect of a high temperature lies in speeding up the reaction.

Finally, sample D shows the most notable improvement of the conductivity and ferromagnetic properties. In addition to this, sample D has the largest metal content among all investigated samples and, most importantly, according to STEM-EELS consists of pure Co in a 20\,nm thick top layer.

A concluding remark is in order concerning the reproducibility of the reported results. In total, four different series of samples, which underwent purification protocols A-D in Fig.~\ref{fChart}, have been studied in the course of preparing this work. The results are reproducible, that attests to the robustness of the elaborated approaches. We therefore believe that the reported purification procedures will find a broad implementation in technology and various fields of research, such as mesoscopic physics, micromagnetism, spin-dependent transport and electronic correlations.
\section{Conclusion}\label{sConclusion}
In summary, we report a comparative analysis of different in-situ, post-growth cleaning approaches to obtain pure Co from Co-FEBID nanostructures. The purification procedure lies in the exposure of heated samples to a H$_2$ atmosphere in conjunction with the irradiation by low-energy electrons. Specifically, the method relies upon the following effects:
\begin{description}
  \item[(i)] Electron-assisted transformation of the amorphous carbon matrix.
  \item[(ii)] Annealing-assisted microstructural modifications of the cobalt clusters.
  \item[(iii)] Hydrogen-assisted removal of carbon due to the catalytic activity of cobalt.
\end{description}
As the main result of this work we state that the combination of annealing at 300\,$^{\circ}$C, H$_2$ exposure and electron irradiation leads to compact, carbon- and oxygen free Co layers down to a thickness of about 20\,nm starting from Co-FEBID structures. These purified structures exhibit metallic conductance properties and behave in magneto-transport measurements like polycrystalline Co thin films. If the initial Co-FEBID structure thickness is adapted to the observed purification depth and moderate thickness shrinkage, direct-write, clean ferromagnetic Co nanostructures with a thickness up to 20\,nm can be prepared.
\ack
HP thanks Prof. Ferdinand Hofer, Prof. Werner Grogger, Prof. Gerald Kotleitner, and Martina Dienstleder for support. HP also acknowledges financial support by the EU FP7 programme (FP7/2007-2013) under grant agreement no. 312483 (ESTEEM2). This work was in parts supported by the Beilstein Institut, Frankfurt/M, within the research collaboration NanoBiC, and was done within the framework of the COST Action CELINA (CM1301).


\providecommand{\newblock}{}

\end{document}